\documentclass[aps,prd,twocolumn,groupedaddress,nofootinbib]{revtex4}

\usepackage{graphicx}
\usepackage{amsmath,amssymb,amsfonts}
\usepackage[colorlinks,citecolor=blue,linkcolor=blue,urlcolor=blue]{hyperref}
\usepackage[applemac]{inputenc}

\usepackage{physics} 
\usepackage[percent]{overpic} 

\newcommand{\be}{\begin{equation}}
\newcommand{\ee}{\end{equation}}
\usepackage[english]{babel}

\usepackage[autostyle, english = british]{csquotes}
\MakeOuterQuote{"}

\begin{document}

\title{Memory and entropy}

\author{
Carlo Rovelli}
\affiliation{Aix Marseille University, Universit\'e de Toulon, CNRS, CPT, 13288 Marseille, France.\\ Perimeter Institute, 31 Caroline Street North, Waterloo, Ontario, Canada, N2L 2Y5.\\ The Rotman Institute of Philosophy, 1151 Richmond St.~N London, Ontario, Canada, N6A 5B7.}

\date{\small\today}

\begin{abstract} \noindent  I study the physical nature of traces (or memories). Surprisingly, (i) systems separation with (ii) temperature differences  and (iii) long thermalization times, are sufficient conditions to produce macroscopic traces. Traces of the past are ubiquitous because these conditions are largely satisfied in our universe.  I quantify these thermodynamical conditions for memory and derive an expression for the maximum amount of information stored in such memories, as a function of the relevant thermodynamical parameters. This mechanism transforms low entropy into available information. 
 \end{abstract}

\maketitle

\section{\bf The problem} 

 The present abounds with traces of the past (footsteps in the sand, craters on the moon, geological strata, photos of us younger,...) and no similar traces of the future. We remember the past, not the future, and this might even be   the very source of the psychological arrow of time \cite{Wolpert, libro}. What is the physical mechanism giving rise to this time asymmetry and, especially, to the great abundance of traces of the past in nature?  
 
The second law of thermodynamics is the only "fundamental" law (including in quantum physics \cite{Einstein,io}) that breaks time-reversal invariance, hence traces are likely to be macroscopic phenomena related to an entropy gradient and ultimately to the past hypothesis \cite{Albert} of primordial low entropy.  Why does past low-entropy yield the ubiquity of traces of the past, and how?  

Building on \cite{Reichenbach, Albert},  I show here that the combined presence of (i) systems separation, (ii) past low entropy, in the form of a temperature difference between systems, and (iii) long thermalization times, is a sufficient condition to produce traces of the past.  

Using this result, I derive an expression for the maximum amount of information stored in memory in this way, as a function of the relevant thermodynamical parameters. 

The problem discussed here is not how to characterise the physical meaning of "memory" in general.   It is to understand why there are so many traces of the past around us, assuming nobody planted them on purpose. More precisely: what is the general thermodynamical mechanism, which appears to be in place in the universe, that generates the abundance of traces we see?   

The relationship between memory and the second law, is studied form a different perspective also in \cite{Wolpert}.

\section{\bf A concrete model for memory}

I use the expressions "memory" or "trace" as synonyms, to indicate a feature of the present configuration of the world that we promptly identify as witnessing a specific particular past event. Examples are photos, written texts, footsteps, impact craters, memories in brains and computers, fossils, gravitational waves emitted from a black hole merger, this very article you are reading, and so on.    There are no analogous traces of future events, in our experience.  To pinpoint the nature of these memories, I describe a simple paradigmatic model.

Consider two physical systems that interact occasionally.  The first, say, is formed by balls moving freely in a closed room, bouncing elastically on the walls and on each other. Assume that the balls are sufficiently heavy and elastic to make any dissipation negligible and that have been bouncing long enough to average out their energy.   The second is formed by several pendulums hanging on ropes from the ceiling. For these, assume that friction in the rope and with air is non-negligible.   Hence oscillations of the pendulums are damped.  The two systems interact  when one of the balls happens to hit elastically one of the pendulums.  Assume that conditions are such that collisions between the balls and the pendulums happen but are rare, during the time span considered. 

\begin{figure}[b]
\includegraphics[width=5cm]{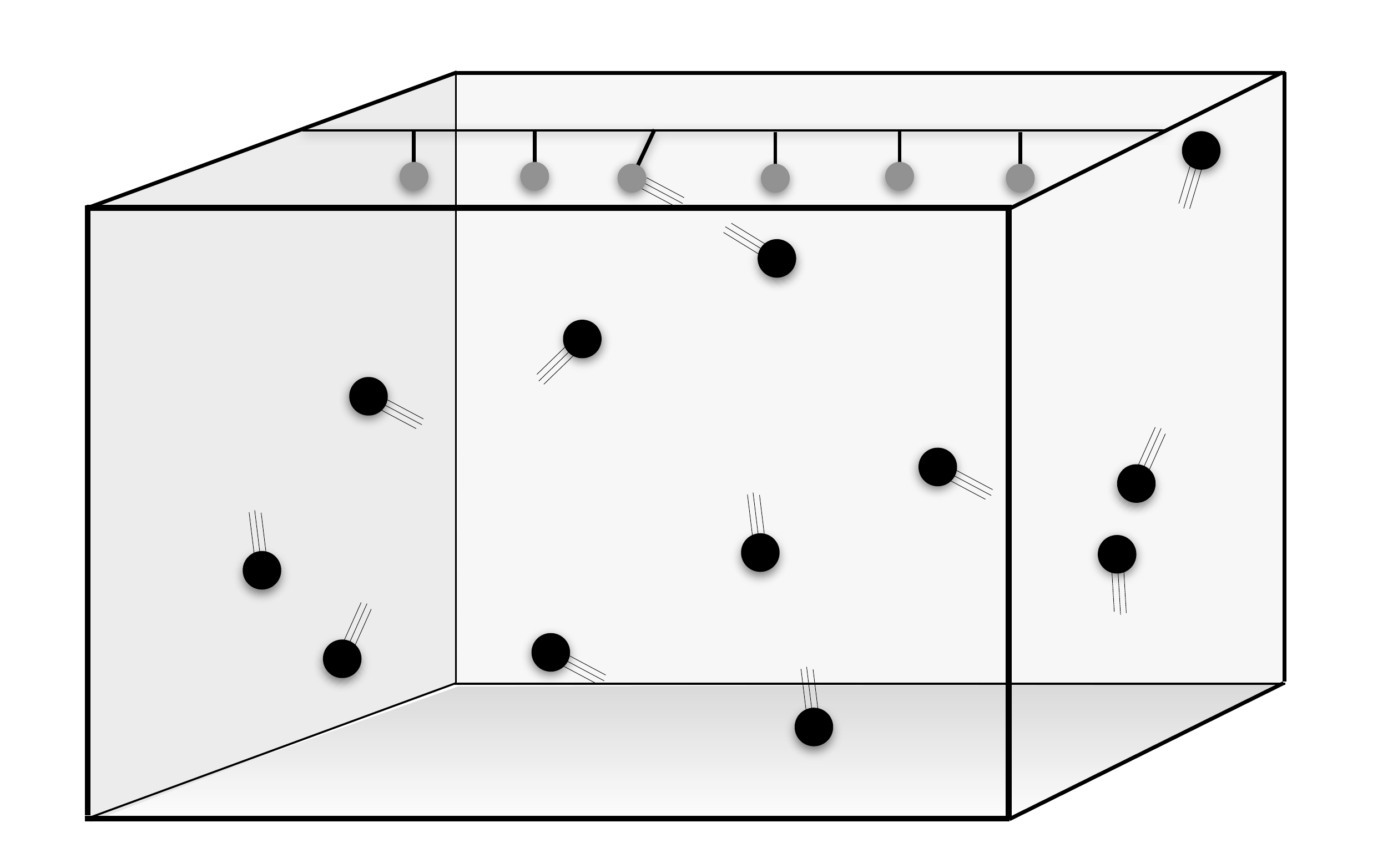}
\caption{\em The model for memory.  The cold damped pendulums are in grey, the hot balls in black. Here the wide oscillation of a single oscillating pendulum is a trace of a past interaction between a ball and a pendulum.}
\end{figure}

Call "event" a collision between a ball and a pendulum.    Consider histories where the pendulums are initially near rest, undergoing small thermal fluctuations, while the balls move fast.   Anytime a ball hits a pendulum, it sets it in motion, the pendulum oscillates for a while. Its oscillations are slowly damped by friction.  

Consider the state of this system at some time $t$.  See Figure 1. Say we see most pendulums near rest, except for a few that oscillate widely.  From these oscillations, we infer that those pendulums were hit by a ball \emph{in the past}.  This is an example of "trace", or "memory".  The memory, namely the oscillation, lasts for a while.  Long after the hit, the excited pendulums are back to near rest, as they were initially.  The state of the pendulums at time $t$ has information about events (interactions between balls and pendulums) in the past, but not in the future.  Something has broken time reversal invariance.  What, and how? 

The first point to observe is that the description above is macroscopic.  By this I mean that the only relevant information given is that "the balls move fast", "the pendulums are near rest". This is \emph{averaged} information.   The point is that it is sufficient to detect a trace.  

A trace is an irreversible phenomenon. This can be seen here by running the history backward in time: one of the few wildly oscillating pendulums collides with ball that happens to absorb its energy nearly entirely.  This looks very implausible, namely highly improbable.  If the phenomenon is irreversible, it needs an initial low entropy, to break time reversal invariance.   Where is the initial low entropy? 

The answer is that the initial low entropy is in the temperature difference between the two systems (balls and pendulums) at the past end of the time interval considered. This in fact decreases slowly at each event.  Indeed, let $E_e$ ("\emph{e}" for environment) be the average energy per degree of freedom of the balls. This defines a temperature $T_e$ via $E_e=\frac12 k T_e$. Here $k$ is the Boltzmann constant.  Let $E_m$ ("\emph{m}" for memory) be the average energy per degree of freedom of the pendulums. This defines a temperature $T_m$ via $E_m=\frac12 k T_m$.   If the two systems had the same temperature, namely if $T_e=T_m$, all degrees of freedom would have the same average kinetic energy, the transfer of energy at each ball-pendulum collision would be equally probable from the balls to the pendulums or viceversa, and there would be no memory.   

To have memory, we need the balls to have more average energy than the pendulums. That is, we need $T_e>T_m$.  The initial difference of temperature is needed here to have memory.  This difference of temperature is a past low entropy condition, because maximal entropy requires $T_e=T_m$.  

At each collision the entropy grows because part of the energy is passed from the balls to the pendulums and the temperature difference decreases. If a small amount of energy $\Delta E$ is transmitted to a pendulum in a collision, this gives rise (in due time after the energy of the pendulum is dissipated) to an entropy increase 
\be
\Delta S\sim \frac{\Delta E}{T_m}-\frac{\Delta E}{T_e}.  \label{S}
\ee
because in thermodynamical terms the transferred energy here behaves like heat transferred.   In a time reversed history, a pendulum decreasing its energy by transmitting it to a ball is precisely like heat going from a cold to a hot body, violating the second law because improbable, although mechanically possible. 

To have memory, we also need $T_e>T_m$ to persist.  Namely, we need the interactions between balls and pendulums to be sparse.  Thus, we need the coupling between the two systems to be sufficiently weak to hold them from converging too rapidly to equilibrium. That is, the thermalization time $\tau_{em}$ of the coupled system must be long, on the scale of the observation time $t_{tot}$.   

We also need the damping of the oscillations of the pendulums to be sufficiently slow to hold the memory.  That is, the thermalisation time $\tau_m$ for the pendulum systems must be sufficiently long on the scale of the time $t_m$ we want the memory to last.    In other words, we need the pendulums themselves to be sufficiently weakly coupled to hold them from converging rapidly to equilibrium.  The memory, or trace, is then neatly identified as a configuration (of a pendulum) which is out of equilibrium (with rope, air and other pendulums), and remains so for a while thanks to the long thermalization time $\tau_m$.  

Notice that the initial temperature difference breaks time reversal invariance. The set of dynamical histories compatible with it is much smaller than the set of histories compatible with the smaller temperature difference at a later time. 

\section{\bf The nature of memory} 

The model discussed above points to a set of simple ingredients sufficient to lead to traces: 
\begin{itemize}
\item[(a)] System separation.  In the example: the balls and the pendulums.  Similarly: the meteorites population and the moon's surface, men walking and the sand, people and the film of the camera.  The separation must be sufficient to permit that thermalization is avoided for a time lapse long with respect to the memory time.  That is, the thermalization time $\tau_{em}$ of the composite system must be much longer than the time lapse considered $t_{tot}$. 

 \item[(b)] Past thermodynamical imbalance, for eaxample a temperature difference.  For instance, there should be higher average energy (hence temperature, $T_e$) in one the two systems (balls, meteorites, walking people, colliding black holes) than in the other $T_m$. Denote "environment" the systems with higher temperature and "memory" the system with lower temperature. 

\item[(c)] Long thermalization time in the  memory system.   The pendulums should not thermalise too fast for memory to hold. The sand should not be equalised by the wind too fast, the photo should not fade too fast, the gravitational waves should not sink in the background radiation,...   That is, the thermalization time of the memory system $\tau_m$ must be longer than the expected duration $t_m$ of the memory. 
\end{itemize}

These ingredients are sufficient to give raise to the phenomena that we recognise as traces, or memories, without any need of anything else.  Traces are the temporary (but long lasting, because of $\tau_m)$ out of equilibrium configurations of the "memory" system, that \emph{follow} the occasional interactions between environment and memory.  They are in the \emph{future} of the interaction because of the time orientation sourced by the initial low entropy.  

That is: let's have two systems, that we call environment and memory, at temperature $T_e$ and $T_m$ respectively.  Let's  $\tau_{em}$ be the thermalization time for the coupled system environment+memory and $\tau_m$ the thermalization time in the memory system alone.  For a total time lapse $t_{tot}$ we have memories lasting a time of order $t_m$ if 
\begin{itemize}
\item[(a)] $\tau_{em}> t_{tot}$,
\item[(b)] $T_e > T_m$,
\item[(c)] $\tau_m > t_m$.
\end{itemize}
It is remarkable that such simple conditions are sufficient to generate traces.   

These conditions are obviously largely realized in our universe, where many subsystems interact weakly, and thermalization is often very slow, giving rise to very common metastable states.  

The scheme described here can probably be generalised, for instance to cover chemical potential, or pressure disequilibria, but it seems to me to capture the core of the reason for the abundance of traces in the universe, and their time orientation.

\section{\bf Information in memory} 

Let the heat capacity of the memory system be $C_m$, and let us take the heat capacity of the environment to be infinite for simplicity (namely let us assume that the environment is large). Then as time passes the temperature of the memory raises.  This can be simply modelled over long enough time scales as
\be
T_m(t) = T_e-(T_e-T_m) e^{-t/\tau_{em}}. \label{Tm}
\ee
Individual memories are lost after a time $\sim\!\tau_m$. 
The total energy transmitted during a time lapse $\tau_{m}$ is of the order of 
\be
\Delta E=C_m\, (T_m(\tau_m)-T_m). 
\ee
This is therefore the average energy dropped into the memory system and not yet thermalised, at any given time. 
With \eqref{Tm}, this gives
\be
\Delta E= C_m (T_e-T_m)(1-e^{-\tau_m/\tau_{em}}).
\ee
Using \eqref{S}, this energy gives rise to an entropy change
\be
\Delta S=  \frac{C_m(T_e-T_m)^2}{T_eT_m}(1-e^{-\tau_m/\tau_{em}}). 
\label{DeltaS}
\ee
This is the entropy increase from which the information stored in the memory is sourced, under the conditions given.  

Memory contains information. This information must come from somewhere.  The only possible source is the entropy increase, since entropy increase is information loss.  Therefore \eqref{DeltaS} determines the maximal amount of information $I$ that can be stored in memory in this way: 
\be
I <\Delta S/k = \frac{C_m(T_e-T_m)^2}{kT_eT_m}(1-e^{-\tau_m/\tau_{em}}). 
\ee

Notice that this is a mechanism that transforms initial low entropy (free energy) into available information.  

Perhaps this transformation of past low entropy into available information plays a role in  phenomena such as the biosphere, where available information plays a huge role. 

In our universe, system separation and very long thermalization times ($\tau_{em}$) are very common.  By far the dominant one is the lack or thermodynamical equilibrium between helium and hydrogen.  Helium and hydrogen  ceased to be in thermal equilibrium since nucleosynthesis, because the rapid cosmological expansion (matter was out of equilibrium with the scale factor \cite{pastlowentropy}) lowered the temperature to a point where the thermalization time became much longer than cosmological times. Helium and hydrogen  have remained out of equilibrium since and this disequilibrium is currently the main source of free energy in the universe \cite{pastlowentropy}.  Occasionally,  energy is dumped from hydrogen to helium: this happens when a star form and burns.   These events are those fuelling all the free energy that nurtures the biosphere. They are a perfect realisation of the mechanism described here and they, indeed, leave very abundant traces of themselves and of their products.

I have long being puzzled by how a disorganised universe, where in the far past matter was in thermal equilibrium,    could have spontaneously evolved into the abundance of informative traces of the past that we see around us. It seems to me that the mechanism described here provides the answer. \\

\centerline{***}

Sincere thanks to Jenann Ismael, David Albert and Andrea di Biagio for reactions, very good criticisms to the first draft of this note, and comments.  This work was made possible through the support of the  FQXi  Grant  FQXi-RFP-1818 and by the QISS grant ID\# 60609 of the John Templeton Foundation.

\end{document}